\documentclass[8.5pt,twoside,twocolumn]{article}
\usepackage[super,sort&compress,comma]{natbib} 
\usepackage{times,mathptm}
\usepackage{sectsty}
\usepackage{balance} 

\usepackage{graphicx} %eps figures can be used instead
\usepackage{lastpage}
\usepackage[format=plain,justification=raggedright,singlelinecheck=false,font=small,labelfont=bf,labelsep=space]{caption} 
\usepackage{fancyhdr}
\newcommand{\oeno}{\oe{}nologie}

\begin{document}

\thispagestyle{plain}
\fancypagestyle{plain}{
%\fancyhead[L]{\includegraphics[height=8pt]{headers/LH.pdf}}
%\fancyhead[C]{\hspace{-1cm}\includegraphics[height=20pt]{headers/CH.pdf}}
%\fancyhead[R]{\includegraphics[height=10pt]{headers/RH.pdf}\vspace{-0.2cm}}
\renewcommand{\headrulewidth}{1pt}}
\renewcommand{\thefootnote}{\fnsymbol{footnote}}
\renewcommand\footnoterule{\vspace*{1pt}% 
\hrule width 3.4in height 0.4pt \vspace*{5pt}} 
\setcounter{secnumdepth}{5}

\makeatletter 
\def\subsubsection{\@startsection{subsubsection}{3}{10pt}{-1.25ex plus -1ex minus -.1ex}{0ex plus 0ex}{\normalsize\bf}} 
\def\paragraph{\@startsection{paragraph}{4}{10pt}{-1.25ex plus -1ex minus -.1ex}{0ex plus 0ex}{\normalsize\textit}} 
\renewcommand\@biblabel[1]{#1}            
\renewcommand\@makefntext[1]% 
{\noindent\makebox[0pt][r]{\@thefnmark\,}#1}
\makeatother 
\renewcommand{\figurename}{\small{Fig.}~}
\sectionfont{\large}
\subsectionfont{\normalsize} 

\fancyfoot{}
%\fancyfoot[LO,RE]{\vspace{-7pt}\includegraphics[height=9pt]{headers/LF.pdf}}
%\fancyfoot[CO]{\vspace{-7.2pt}\hspace{12.2cm}\includegraphics{headers/RF.pdf}}
%\fancyfoot[CE]{\vspace{-7.5pt}\hspace{-13.5cm}\includegraphics{headers/RF.pdf}}
\fancyfoot[RO]{\footnotesize{\sffamily{1--\pageref{LastPage} ~\textbar  \hspace{2pt}\thepage}}}
\fancyfoot[LE]{\footnotesize{\sffamily{\thepage~\textbar\hspace{3.45cm} 1--\pageref{LastPage}}}}
\fancyhead{}
\renewcommand{\headrulewidth}{1pt} 
\renewcommand{\footrulewidth}{1pt}
\setlength{\arrayrulewidth}{1pt}
\setlength{\columnsep}{6.5mm}
\setlength\bibsep{1pt}

\twocolumn[
  \begin{@twocolumnfalse}
\noindent\LARGE{\textbf{Micelle formation, gelation and phase separation of amphiphilic multiblock copolymers$^\dag$}}
\vspace{0.6cm}

\noindent\large{\textbf{Virginie Hugouvieux,$^{\ast}$\textit{$^{a,d,e}$} Monique A. V. Axelos,\textit{$^{b}$} and
Max Kolb\textit{$^{c}$}}}\vspace{0.5cm}

\noindent\textit{\small{\textbf{Received Xth XXXXXXXXXX 20XX, Accepted Xth XXXXXXXXX 20XX\newline
First published on the web Xth XXXXXXXXXX 200X}}}

\noindent \textbf{\small{DOI: 10.1039/b000000x}}
\vspace{0.6cm}

\noindent \normalsize{The phase behaviour of amphiphilic multiblock copolymers with a large number of blocks in semidilute solutions is studied by lattice Monte Carlo simulations. The influence on the resulting structures of the concentration, the solvent quality and the ratio of hydrophobic to hydrophilic monomers in the chains has been assessed explicitely. Several distinct regimes are put in evidence. For poorly substituted (mainly hydrophilic) copolymers formation of micelles is observed, either isolated or connected by the hydrophilic moieties, depending on concentration and chain length. For more highly substituted chains larger tubular hydrophobic structures appear which, at higher concentration, join to form extended hydrophobic cores. For both substitution ratios gelation is observed, but with a very different gel network structure. For the poorly substituted chains the gel consists of micelles cross-linked by hydrophilic blocks whereas for the highly substituted copolymers the extended hydrophobic cores form the gelling network. The interplay between gelation and phase separation clearly appears in the phase diagram. In particular, for poorly substituted copolymers and in a narrow concentration range, we observe a sol-gel transition followed by an inverse gel-sol transition when increasing the interaction energy. The simulation results are discussed in the context of the experimentally observed phase properties of methylcellulose, a hydrophobically substituted polysaccharide.}
\vspace{0.5cm}
 \end{@twocolumnfalse}
  ]

\section{Introduction}

\footnotetext{\textit{$^{a}$~INRA, UMR1083 Sciences pour l'\oeno, F-34060 Montpellier, France. Fax: +33 (0)499612857; Tel: +33 (0)499612758; E-mail: Virginie.Hugouvieux@supagro.inra.fr}}
\footnotetext{\textit{$^{b}$~INRA, UR1268 Biopolym\`eres Interactions Assemblages, F-44300 Nantes, France. }}
\footnotetext{\textit{$^{c}$~Laboratoire de Chimie, Ecole Normale Sup\'erieure de Lyon, 46 all\'ee d'Italie, 69364 Lyon Cedex 07, France. }}
\footnotetext{\textit{$^{d}$~Montpellier SupAgro, UMR1083 Sciences pour l'\oeno, F-34060 Montpellier, France.}}
\footnotetext{\textit{$^{e}$~Universit\'e Montpellier 1, UMR1083 Sciences pour l'\oeno, F-34060 Montpellier, France.}}

Amphiphilic block copolymers\cite{lindman_amphiphilic_2000} are particularly versatile macromolecules because they allow for a rich variety of different structures. They may be built from many different kinds of monomers with different chemical properties. Their length and the number of blocks of each species can be tuned at will, from di- and triblock to multiblock copolymers\cite{lindman_amphiphilic_2000}. Their architectures can be linear, branched or star-like, the blocks may be distributed randomly or regularly. Also, copolymers may interact through hydrogen bonding, hydrophobic interactions or electrostatic forces, thus resulting in different, possibly competing, types of behaviour in solution, depending on the physico-chemical parameters such as temperature, pH, ionic strength and type of solvent\cite{chandler_interfaces_2005,hummer_hydrophobic_1998,meyer_recent_2006}. Due to their dual affinity for polar and apolar solvents these macromolecules can self-assemble in order to minimize their free energy. This gives rise to a wide range of structures and properties, from swollen copolymers in a good solvent to microphase formation, gelation and phase separation in a poor solvent. Amphiphilic block copolymers have attracted a great deal of attention from the biomedical field\cite{bae_biodegradable_2000}, with applications such as drug delivery vectors\cite{kataoka_block_2001}, nanoparticle stabilizers, nanoreservoirs, emulsion stabilizers, wetting agents, rheology modifiers\cite{riess_micellization_2003,bhatia_block_2001} or as injectable scaffold materials for tissue engineering\cite{fatimi_rheological_2008}. 

In the present work our focus is on the specific case of strictly alternating linear multiblock copolymers consisting of two kinds of monomers, hydrophobic (H) and polar (P). This may be viewed as a generic, simplified model for the understanding of the structural aspects and the phase behaviour of hydrophobically substituted polysaccharides \cite{sarkar1979thermal}, an important example of which is methylcellulose. Methylcellulose is obtained by methylation of some hydroxyl groups of the cellulose polymer. It results in alternating blocks of more or less hydrophobic monomers, whose proportions and lengths depend on the degree of methylation of each glucose cycle (from 0, hydrophilic to 3, hydrophobic). There is experimental evidence that the substituted monomers are indeed arranged in a block pattern\cite{hirrien_thermogelation_1998,kobayashi_thermoreversible_1999,arisz_substituent_1995}. Hence we expect that the simulation of a hydrophobic/polar multiblock copolymer can give insight into the structure and properties of methylcellulose as a function of thermodynamic conditions\cite{chevillard_phase_1997,guillot_non-self-similar_2000}.

There is a vast body of theoretical and experimental studies on diblock and triblock copolymers\cite{zhang_multiple_1995,riess_micellization_2003,binder_monte_2000,floriano_micellization_1999,li_kinetics_2010,li_effect_2005,wijmans_simulation_2004}. Much fewer studies deal with the simulation of amphiphilic multiblock copolymers, and most of them address the questions of their structure and kinetics in dilute solutions where only intramolecular interactions play a role\cite{dasmahapatra_pathway_2007,chushak_coarse-grained_2005}. It was convincingly demonstrated that multiblock copolymers not only form intramolecular chains of micelles\cite{tanaka_intramolecular_2000, cooke_collapse_2003} but also tubular structures and intramolecular bilayers\cite{hugouvieux_amphiphilic_2009}. The formation of this variety of intramolecular structures is mainly influenced by two parameters, the effective interaction strength between the H monomers and the hydrophobic substitution ratio. It is a consequence of the balance between the entropic penalty of constraining the polar P monomers and the energetic gain due to the self-assembly of the hydrophobic H monomers. The formation of intramolecular pearl-necklaces of micelles in multiblock copolymers was justified theoretically by Halperin\cite{halperin_collapse_1991}. For the case of amphiphilic graft copolymers, the existence of intramolecular pearl-necklaces of micelles and cylindrical micelles was argued theoretically and directly observed in simulation studies\cite{borisov_amphiphilic_2005,kosovan_amphiphilic_2009}. Also, it was shown experimentally that amphiphilic graft copolymers in dilute solutions form either unimolecular rodlike chains of micelles or unimolecular rodlike micelles, depending on the quality of the solvent\cite{kikuchi_unimolecular_1996,kikuchi_unimolecular-micelle_1996}.

Regarding semidilute solutions, where both intra- and intermolecular interactions are present, Anderson and Travesset\cite{anderson_coarse-grained_2006} performed simulations of the phase behaviour of pentablock copolymers and found different regimes as a function of concentration and solvent quality of the hydrophilic blocks: a swollen gel, giant micelles, lamellar and lamellar-perforated phases were observed. Glotzer \textit{et al.}\cite{glotzer_physical_1993} showed in a two-dimensional model calculation that multiblock copolymers, unlike diblock copolymers, micro-phase separate and form a gel network. Gindy \textit{et al.}\cite{gindy_phase_2008} focussed on the role of the number $m$ and length $n$ of the blocks in (H$_n$P$_n$)$_m$ multiblock copolymers containing the same number of H and P monomers. They showed that these copolymers may either evolve by micro-phase separation or by macroscopic phase separation, depending on the $m/n$ ratio. From the experimental side microphase separation of graft copolymers was studied as a function of the hydrophilic and hydrophobic block lengths\cite{hourdet_small-angle_1998}. From a rheological point of view Regalado \textit{et al.}\cite{Regalado_viscoelastic_1999} advanced evidence for three viscoelastic regimes in semidilute solutions of multiblock copolymers, depending on concentration, hydrophobic ratio and hydrophobic block length. A study of the impact of pH and temperature on the supramolecular self-assembly of pentablock copolymers in solution\cite{determan_supramolecular_2006,determan_ph-_2008} showed that increasing the pH leads to an increase in the hydrophobicity which, in turn, triggers the formation of spherical micelles which eventually become cylindrical (also called wormlike). Lintuvuori and Wilson\cite{lintuvuori_coarse-grained_2009} focussed on the behaviour of rod-coil multiblock copolymer melts when cooling a system through the order disorder transition (ODT). They gave evidence for the formation of micellar, nematic, lamellar and gyroid phases and showed that the length of the rods influences the temperature of the ODT. Recently, the influence of a block profile that varies along the copolymer has been studied. The resulting micelles significantly deviate from the ones obtained for diblock copolymers\cite{kuldova2010self}. The dynamics of block copolymer systems has been critically discussed by \citet{denkova2010non}, paying particular attention to the consequences of slow dynamics on experiments and simulations.

In the present work we focus on the behaviour of linear multiblock copolymers in semidilute solutions, where both intra- and intermolecular interactions come into play. In contrast with previous investigations, in our model each copolymer is made of a large number of hydrophobic and hydrophilic blocks (up to 15 of each).

The role of the polymer concentration and the strength of the hydrophobic effect on the phase behaviour is studied. Furthermore, the role of the substitution ratio is investigated. Different patterns are identified (including spherical and cylindrical micelles, gel formation and phase separation) and the transitions between these different structures are assessed.

The paper is organized as follows: Section \ref{sec:model} describes the simulation method, the model and the simulation details. The properties computed for gaining an understanding of the behaviour of the system are defined in Sec. \ref{sec:prop}. The results are presented in Sec. \ref{sec:results} for two representative substitution ratios. They are discussed and compared with previous experimental and simulation studies in Sec. \ref{sec:discussion}.

\section{Simulation model and method} \label{sec:model}

\subsection{Simulation method} In the present work an efficient lattice MC model is used, which we introduced in detail in a previous article\cite{hugouvieux_amphiphilic_2009}. Polymer chains are discretized on a face-centered cubic lattice, thus allowing for a large number of nearest neighbour sites and bond angles. Periodic boundary conditions are imposed. A lattice site may be occupied by zero, one or two monomers. Adjacent monomers in a chain occupy the same site or nearest neighbour sites on the lattice (double occupancy is restricted to neighbouring monomers along a polymer, see \citet{hugouvieux_amphiphilic_2009} for full details). Thus a lattice site represents a volume of two monomers. The excluded-volume constraint is strictly respected by all monomers and the polymers are never allowed to cross each other. Dynamics is implemented through attempted nearest neighbour hops of randomly selected monomers. Moves are accepted if the final configuration again respects bond length and excluded volume restrictions. For the interacting hydrophobic monomers a trial move furthermore has to satisfy the Metropolis criterion to be accepted\cite{hugouvieux_amphiphilic_2009}.

\subsection{Model} We consider semidilute solutions of linear, regularly alternating multiblock copolymers made of two kinds of monomers, hydrophobic (H) or hydrophilic/polar (P). The solvent is not dealt with explicitly, and the hydrophobic effect is mimicked by an effective nearest-neighbour attractive interaction between the H monomers, which implicitely expells the polar solvent from the vicinity of the hydrophobic monomers. This attractive force acts between any two H monomers that occupy the same or two nearest-neighbour sites (called interacting monomers in the following) and is given by the interaction energy $E_i < 0$, expressed in units of the thermal energy $k_B T$. Increasing the absolute value of $E_i$ corresponds to a strengthening of the H-H interaction and hence is equivalent to a lowering of the temperature (except for the case of systems with a lower critical solution temperature (LCST), for which it corresponds to an increase in temperature\cite{gil2004stimuli}). No difference is made between intra- and intermolecular interactions between H monomers. The P monomers neither interact explicitely with the H or the P monomers nor with the solvent. The pattern of the blocks in the chains is described as $(H_{B_H}P_{B_P})_n$, with $B_H$ the length of a hydrophobic block, $B_P$ the length of a polar block and $n$ the number of times such a diblock is repeated in the chain. In this study $n$ takes values between 5 (10 blocks) and 15 (30 blocks). Hence the focus is on systems of copolymers with a large number of blocks. The degree of polymerization (total number of monomers) of the chain is $N_m = n \times (B_H+B_P)$. The hydrophobic substitution ratio is defined as $P_{sub} = \frac{B_H}{B_H+B_P}$. Distances (wave vectors) are expressed in units of lattice spacing ($2 \pi$ over lattice spacing).

\subsection{Simulation details} Simulations were performed for different values of concentration, solvent quality, substitution ratio and copolymer chain length. The concentration $\phi$ is defined as the fraction of the total available volume which is occupied by monomers. Because of the double occupancy the total volume $V=L^3$ is twice the number of available lattice sites $N_s=\frac{1}{2}L^3$, where $L$ is the periodicity. Simulations were performed on monodisperse many-chain systems with chain lengths $N_m=100, 300$, for different concentrations $\phi$ ranging from $0.005$ to $0.24$, and for different interaction energies $E_i$ from -0.3 to -0.6. This enables a broad range of structural features to be encompassed, from dilute good solvent conditions to concentrated solutions of strongly interacting copolymers. Also, different substitution ratios $P_{sub}$, chain lengths $N_m$ and numbers of polymer $N_p$ give insight into the phase behaviour and finite-size effects. The simulations are performed in the canonical $NVT$ ensemble, with fixed number of monomers $N$, volume $V$, and temperature $T$. For each set of conditions, several independent simulation runs are performed (typically 8 to 16), starting from distinct random initial configurations, generated by a random polymerization process. Time is measured in Monte Carlo step (MCS) where one MCS corresponds to $N_m \times N_p$ trial moves. The systems are equilibrated for some $10^7$ MCS. The relaxation of the time auto-correlation function of the end-to-end vector of the polymers is used as a criterion for equilibration. During the production run configurations are analysed periodically (every $2.10^5$ to $10^6$ MCS depending on $E_i$ and $\phi$) in order to compute the properties of the system. Typical production runs range from $10^7$ to $8.10^7$ MCS, depending on the relaxation time of the system. Simulations were performed for two values of the substitution ratio, $P_{sub}=0.2$ and $0.5$, representative of different copolymer architectures. The simulation details are summarized in Table \ref{tab:simul}, the concentration being varied through changes of the system size.\\

\begin{table}[h!]
\small
\caption{Parameters used in the simulations} \label{tab:simul}
\begin{tabular*}{0.5\textwidth}{@{\extracolsep{\fill}}cccccc}
\hline
$P_{sub}$  & $B_H$ & $B_P$ &$N_p$ & $N_m$ & $\phi$\\
\hline
0.2 & 4 & 16 & 60 & 300 & 0.005 - 0.04\\
0.2 & 4 & 16 & 100 & 100 & 0.005 - 0.04\\
\hline
0.5 & 5 & 5 & 50 & 100 & 0.005 - 0.05\\
0.5 & 5 & 5 & 300 & 100 & 0.005 - 0.24\\
\hline
\end{tabular*}
\end{table}

\section{Properties} \label{sec:prop}

Direct visualization of the configurations gives valuable qualitive information about the structure of the system. The calculation of experimentally measurable quantities from the simulations allows for a quantitative assessment of these properties and of the influence of the different physical parameters. The calculated quantities are defined in the following.

\subsection{Clusters} A cluster is defined as a set of interacting copolymers. Two copolymers interact whenever at least one H monomer of the first copolymer is nearest neighbour to an H monomer of the second copolymer. A cluster comprises all the monomers (H and P) of the involved chains. If polymer P$_1$ interacts with P$_2$ and, P$_3$ interacts with P$_2$, all three polymers belong to the same cluster. Thus connectivity through the hydrophilic blocks within a copolymer is part of the cluster definition.

\subsection{Hydrophobic cores} A hydrophobic core is defined as a set of interacting H monomers, ignoring the P monomers and the connectivity of the chains. If the H monomer H$_1$ interacts with H$_2$ and H$_2$ interacts with H$_3$ all three belong to the same core. Note that a cluster may contain several hydrophobic cores. If we consider a single chain, it always belongs to a single cluster for connectivity reasons, but it may contain as many hydrophobic cores as the number of H blocks in a chain, especially in good solvent conditions. We use the following notation to define different useful observables: $N_H^c$ is the number of H monomers in a given core, $N_{ch}^c$ is the number of chains whose H monomers are part of a given core and $N_H^{p,c}$ is the number of H monomers of a given polymer that belong to a given core. All the distributions $W(N)$ presented below are normalized as follows: if $N$ occurs $O(N)$ times, $W(N)=N.O(N) / \sum_{N'} N'.O(N')$ (e.g. $W(N_H^c)$ is the fraction of H monomers in cores of size $N_H^c$).

\subsection{Percolation} A cluster is considered to be percolating if it spans the whole simulation box, i.e. if one or more of its dimensions are larger than the size of the box. A configuration percolates if it contains at least one percolating cluster. For a given set of conditions the probability of percolation, $P_{percolation}$, is defined as the fraction of percolating configurations. We take $P_{percolation} = 0.5$ to be the onset of gelation. 

\subsection{Structure factor} Calculating the scattering intensity of the simulated systems yields quantitative information on structural correlations which may be compared directly with experimental data from scattering experiments\cite{kobayashi_thermoreversible_1999}. Because the structure of the monomers is neglected in our simulations, there is equivalence between the scattering intensity and the structure factor and the two terms will be used interchangeably. The structure factor is defined as: \begin{equation} \label{eq:Sq}
S(q) = \sum_{j=1}^{N} \sum_{k=1}^{N} \left< \exp{[-i\vec{q} \cdot (\vec{r}_j - \vec{r}_k)]} \right>
\end{equation}
where $N$ is the number of monomers in the system, $\vec{q}$ a vector of the reciprocal space and $q$ its norm, $\vec{r}_j$ and $\vec{r}_k$ the positions of monomers $j$ and $k$, respectively.
$S(q)$ gives information about the structural correlations in the system on different length scales. It is useful to compute partial structure factors defined as the separate contributions to $S(q)$ of only the H or only the P monomers. This is eventually to be compared with experiments using labelled chains\cite{lairez2005diffusion}.

\section{Results} \label{sec:results}

It has been shown for isolated copolymers, that the substitution rate determines the pattern developping, through a balance between energetic (hydrophobic) and entropic (hydrophilic) contributions\cite{hugouvieux_amphiphilic_2009}. As will be shown below, there are qualitative differences of the resulting structures for different substitution rates. Therefore we will present data for two distinct substitution rates, each being representative of one scenario. We distinguish between a low substitution rate case, where the screening of the hydrophilic coronae is complete and a high substitution rate case where the hydrophilic screening is incomplete. In general the properties of a copolymer do not only depend on the substitution ratio, but they depend explicitely on both the hydrophilic and hydrophobic block lengths in a non-trivial way.  We have considered different possibilities studying a single copolymer\cite{hugouvieux_amphiphilic_2009}, here we choose one case for each scenario and call it the low and the high substitution ratio case, respectively.

\subsection{Low substitution ratio, $P_{sub}=0.2$}

\begin{figure}[h!]
\centering
 \includegraphics[width=8.3cm]{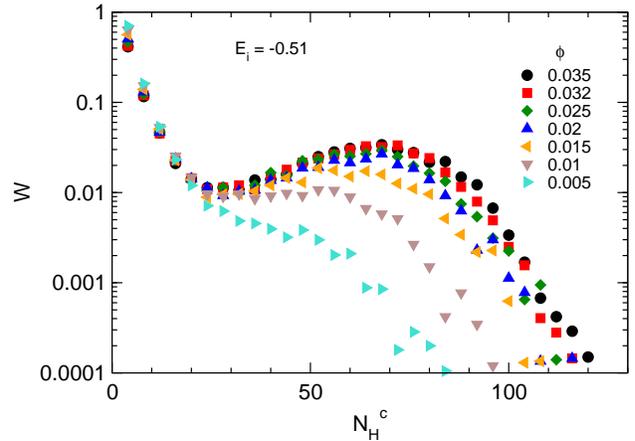}
 \caption{Size distribution of the hydrophobic cores, $W(N_H^c)$, for different concentrations $\phi$ at $E_i=-0.51$, $P_{sub}=0.2$, $B_H=4$, $B_P=16$, $N_p=100$, $N_m=100$.}
 \label{fig:psub02_core_phi}
\end{figure}

\begin{figure}[h!]
\centering
 \includegraphics[width=8.3cm]{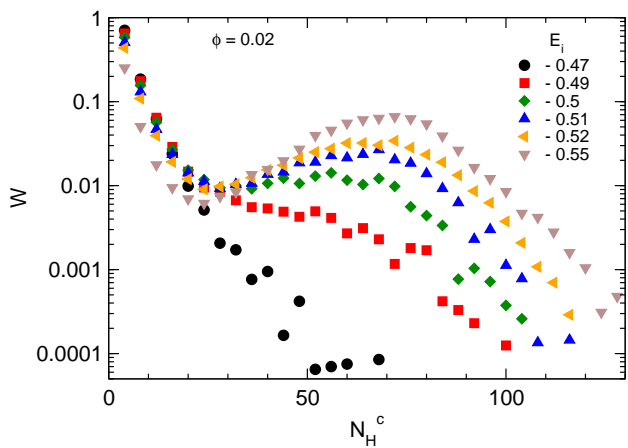}
 \caption{Size distribution of the hydrophobic cores, $W(N_H^c)$, for different interaction energies $E_i$ at $\phi=0.02$, $P_{sub}=0.2$, $B_H=4$, $B_P=16$, $N_p=100$, $N_m=100$.}
 \label{fig:psub02_core_Ei}
\end{figure}

We first present data for a low substitution ratio with an excess of hydrophilic monomers. The influence of the concentration and the interaction energy on the formation of hydrophobic cores were monitored separately. Figures \ref{fig:psub02_core_phi} and \ref{fig:psub02_core_Ei} show the hydrophobic core size distribution either as a function of concentration or interaction energy. The core sizes are expressed in number of H monomers per core, $N_H^c$, which is necessarily a multiple of the number of H monomers in a block, $B_H$.
The core size distribution as a function of concentration is shown in Figure \ref{fig:psub02_core_phi} and has the typical characteristics of a system of surfactant micelles\cite{larson1985mc,rosen2004surfactants}. In the present context a micelle is a structure with an H core surrounded by a P shell. Its shape is either spherical, tubular or lamellar. The appearence of a bump in the size distribution of the hydrophobic cores $N_H^c$ (when the slope at the inflexion point becomes positive) indicates the distinct presence of a well-defined core size, the micelle core. This is used as a definition for the critical micelle concentration (CMC)\cite{israelachvili_intermolecular_1991}. A similar criterion has been used for monitoring the formation of surfactant micelles\cite{wijmans_simulation_2004}. In Figure \ref{fig:psub02_core_phi} , at $E_i=-0.51$, the CMC is $\phi=0.01$ and the distribution of core sizes becomes approximately concentration-independent above $\phi=0.02$, with a stable peak in the micelle size distribution developping around 70 H monomers. Since each copolymer contains 20 H monomers, typical micelles necessarily contain hydrophobic blocks from different polymers, they are multimolecular. A somewhat different behaviour is observed when increasing the interaction strength (see Fig.~\ref{fig:psub02_core_Ei}) at a fixed concentration $\phi=0.02$. There is a systematic increase of the number of large micelles with increasing interaction strength while the maximum of the distribution lies around 65 H monomers and increases slightly with energy.
The number of isolated H blocks (i.e. H blocks that do not interact with any other H block) decreases both with concentration and interaction strength. At fixed $E_i=-0.51$, it decreases by a factor of $\approx 1.7$ upon increasing the concentration from $\phi=0.005$ to $0.035$, and at fixed $\phi=0.02$, it even decreases by a factor of $\approx 2.8$ when increasing the interaction strength from $E_i=-0.47$ to $-0.55$. Similar simulations have been performed for longer chains, $N_m=300$ and $N_p=60$ (see Table \ref{tab:simul}). The size distribution of the hydrophobic cores for the longer copolymers is essentially the same both with regard to its general shape and the value of the most probable core size. Consistent with the observation that the micellar structure is solely stabilised by the balance between the energy of its hydrophobic core and the entropy of the surrounding hydrophilic shell and does not depend on chain length, the multimolecular aspect of the core structure for $N_m=100$ at low concentration becomes essentially monomolecular for $N_m=300$. For the given set of parameters $N_m=300$ happens to be just the limit of the chain length beyond which one expects that a pearl-necklace of micelles starts to form\cite{hugouvieux_amphiphilic_2009}.
\begin{figure}[h!]
\centering
 \includegraphics[width=8.3cm]{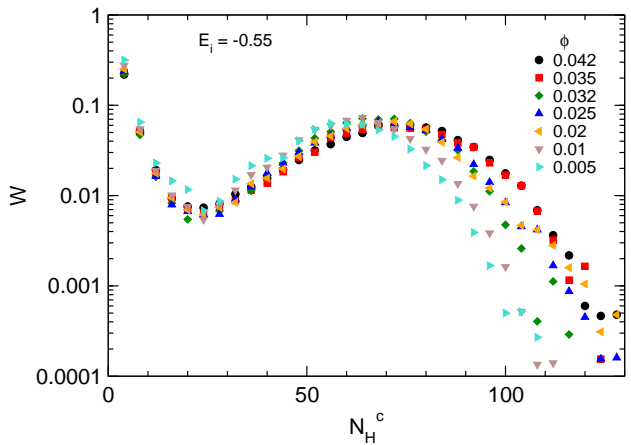}
 \caption{Size distribution of the hydrophobic cores, $W(N_H^c)$, for different concentrations at a stronger interaction energy $E_i=-0.55$, $P_{sub}=0.2$, $B_H=4$, $B_P=16$, $N_p=100$, $N_m=100$. Micellization occurs at any concentration, the CMC is equal to zero.}
 \label{fig:core_size_vs_phi_psub02}
\end{figure}

In Figure \ref{fig:core_size_vs_phi_psub02} we show the evolution of the core size distribution at $P_{sub}=0.2$ for a stronger interaction energy $E_i=-0.55$, for different concentrations. By contrast with $E_i=-0.51$ micelles are now present at any concentration and the core size distribution becomes almost concentration-independent, with a peak between 60 and 70 H monomers. Slight deviations are visible at the lowest concentrations. Hence the difference between the core size distributions at $E_i=-0.51$ and $E_i=-0.55$ is that in the latter case the CMC is equal to zero. This is because, at $E_i=-0.55$, the effective concentration of the H monomers along a single copolymer is high enough such that they always collapse into a micellar structure.
\begin{figure}[h!]
\centering
 \includegraphics[width=8.3cm]{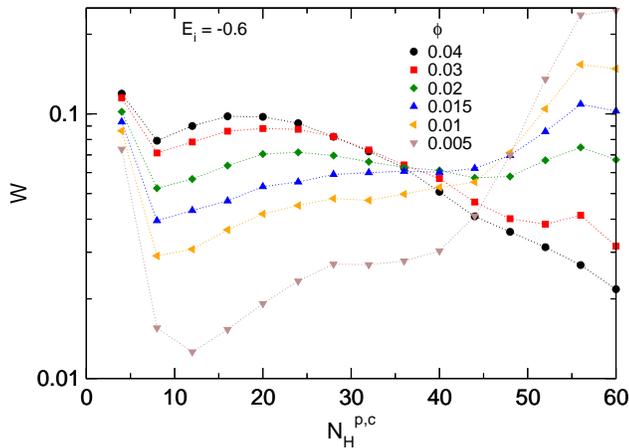}
 \caption{Size distribution of the subset of H monomers of a given copolymer belonging to a given core, $W(N_H^{p,c})$, for different concentrations at $E_i=-0.6$, $P_{sub}=0.2$, $B_H=4$, $B_P=16$, $N_p=60$, $N_m=300$. The maximum value on abscissa corresponds to the total number of H monomers in a copolymer.}
 \label{fig:NH_pc_psub02_Nm300}
\end{figure}
Simulations with $N_p=60$ and $N_m=300$ give very similar results, with the notable exception that the micellar peak is much more narrow at low concentration. With increasing concentration the peak becomes significantly broader (not shown). We attribute this to the fact that the commensurability between the number of H monomers in a copolymer and in a micellar core is better at low concentration. In fact, when increasing concentration, the number of different chains participating in a core increases and the distribution develops a broad maximum at a value close to 3 copolymers per core. Correspondingly, the distribution of $N_H^{p,c}$, the number of H monomers in a core belonging to a given polymer, changes its character completely, as shown in Figure \ref{fig:NH_pc_psub02_Nm300}. At low concentration it is sharply peaked at $N_H^{p,c}=60$, which is just the value when all H monomers of a copolymer belong to the same core. At high concentration there is instead a broad maximum around $N_H^{p,c}=20$. As the size of a core does not change when concentration is increased, the number of different cores connected by the hydrophilic P blocks of a polymer must increase with concentration. This is directly confirmed by the measurements of the radius of gyration of a copolymer, which increases from $R_g^2=19$ at $\phi=0.005$ to $R_g^2=44$ at $\phi=0.041$.

\begin{figure}[h!]
\centering
 \includegraphics[width=8.3cm]{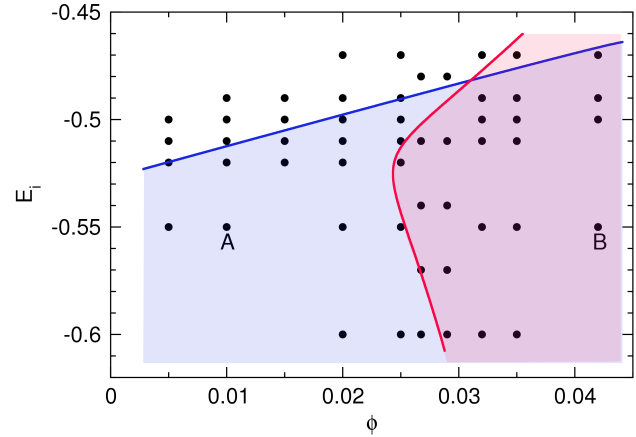}
 \caption{Phase diagram of all systems simulated with short hydrophobic and long hydrophilic blocks: formation of micelles and/or percolation as a function of energy $E_i$ and concentration $\phi$ for $P_{sub}=0.2$, $B_H=4$, $B_P=16$, $N_p=100$, $N_m=100$ and box sizes ranging from $L=62$ to $L=126$. Dots denote the parameter values where simulations were performed. Lines separate the different phases. Formation of micelles is observed in the blue area while gelation occurs in the red area. Note that around $\phi \approx 0.026$ there is successively a sol-gel and then a gel-sol transition upon increasing $|E_i|$. In Fig.~\ref{fig:snapshot_psub02} snapshots of typical configurations are displayed, they correspond to points A and B in Fig.~\ref{fig:diag_psub02} .}
 \label{fig:diag_psub02}
\end{figure}

The phase diagram in Figure \ref{fig:diag_psub02} summarizes the simulations performed at $P_{sub}=0.2$, as a function of concentration and interaction energy. Depending on the parameter values the system may form spherical micelles and/or a gel. The ($E_i$,$\phi$) conditions where the system contains micelles are determined using the inflexion point criterion of the core size distribution (blue line in Fig.~\ref{fig:diag_psub02}). Geometric percolation is taken as the hallmark of gelation\cite{stauffer1994introduction} (red line in Fig.~\ref{fig:diag_psub02}). Four regimes can be distinguished: no micelles and no percolation; only micelles; only percolation; both micelles and percolation. Let us first consider the evolution of the system with respect to micelle formation. For $E_i \le -0.52$ micelles are observed no matter how low the concentration. On the other hand, for $E_i\ge -0.52$, the formation of micelles depends on the concentration of the system as shown in Fig.~\ref{fig:psub02_core_phi}: the lower the interaction strength, the larger the concentration needed for the onset of micelle formation (CMC)\cite{israelachvili_intermolecular_1991}. This relationship is almost linear. 

\begin{figure}[h!]
\centering
 \includegraphics[width=8.3cm]{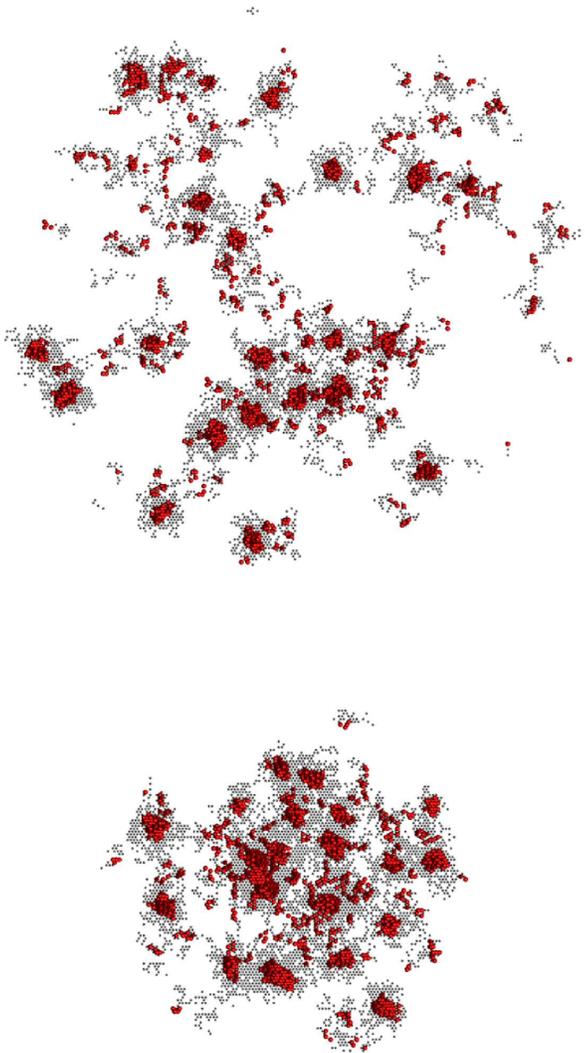}
\caption{Snapshots of typical configurations for $P_{sub}=0.2$, i.e. short hydrophobic and long hydrophilic blocks. Top: micelle phase, point A in Fig.~\ref{fig:diag_psub02} (after $10^7$ MCS). Bottom: gelled micelle phase, point B in Fig.~\ref{fig:diag_psub02} (after $10^7$ MCS). Note that both simulations are performed with the same numbers of monomers but at different concentrations, which explains the difference in box size. Red beads: H monomers. Black dots: P monomers. Some monomers seem isolated because of the periodic boundary conditions.}
 \label{fig:snapshot_psub02}
\end{figure}

A snapshot of a configuration corresponding to the point A in Fig.~\ref{fig:diag_psub02} is shown at the top of Fig.~\ref{fig:snapshot_psub02}. It shows a rather dilute, strongly fluctuating set of hydrophobic cores surrounded by P monomers, the micelles. Percolation sets in for concentrations larger than $\phi=0.025$. The shape of the sol-gel phase boundary is curved such that, in a narrow concentration range, the system first undergoes a sol-gel and then a gel-sol transitions when increasing the interaction strength at constant concentration. This can be understood in view of the results presented in Fig.~\ref{fig:core_size_vs_phi_psub02}. Increasing the interaction strength increases the core size. Because of the compactness of the hydrophobic cores, this leads to an effective shortening of the chain which, in turn, prevents gelation. There is a large domain in which micelle formation and percolation overlap. Point B in Fig.~\ref{fig:diag_psub02} lies in this domain of $E_i$ and $\phi$ values. 
Typical configurations in this zone show a set of micelles consisting of a hydrophobic core surrounded by and connected to each other with blocks of hydrophilic monomers (see Fig.~\ref{fig:snapshot_psub02}, bottom). Complementary results (not shown) obtained for $N_p=60$ and $N_m=300$ have the same phase structure but with the percolation line shifted to lower concentrations. This is consistent with the fact that increasing the chain length facilitates cross-linking.

\begin{figure}[h!]
\centering
 \includegraphics[width=8.3cm]{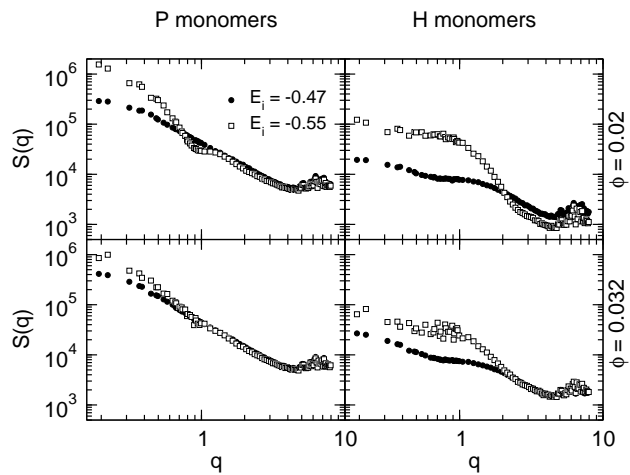}
 \caption{Partial structure factors $S(q)$ for $P_{sub}=0.2$. Top: $\phi=0.02$, bottom: $\phi=0.032$. Left: P monomers only, right: H monomers only. At $E_i=-0.55$, the presence of cores is clearly visible in the H structure factors.}
 \label{fig:sq_psub02}
\end{figure}

Structure factors yield valuable information on the spatial organisation of the system \cite{higgins1997polymers}. For copolymers it is particularly useful to calculate partial structure factors, as they give insight into the spatial correlations of each of the two kinds of monomers separately. This is shown in Fig.~\ref{fig:sq_psub02} for $P_{sub}=0.2$ for two different concentrations $\phi$. The partial structure factors are defined by equation \ref{eq:Sq}, where the summations are now restricted to the P or the H monomers, respectively. At $\phi=0.02$ the strongly interacting system ($E_i=-0.55$) shows the typical features of spherical core-shell structures \cite{forster1998scattering,berndt2006influence} with a Guinier plateau for the H monomers corresponding to the size of the cores and a lack of intensity for the P monomers at the same length scale. Also, a Porod representation of $S_H(q)$ (not shown) exhibits a plateau corresponding to the presence of a sharp interface between the H and P monomers. This contrasts the same analysis for the more weakly interacting system at $E_i=-0.47$ where no micellar core is visible.

At higher concentration, $\phi=0.032$, the structure is less well-defined: the Guinier plateau at $E_i=-0.55$ for the H monomers shows more statistical noise at low $q$ and the decrease in intensity at intermediate $q$ for the P monomers is less pronounced. The Porod representation of $S_H(q)$ no longer shows the plateau characteristic of a sharp segregation of H and P monomers. Concomitant gelation and core formation in this system can explain these observations: the cross-linking of the micelles impedes the formation of well-defined core-shell structures with low interfacial tension. The increase of the large scale statistical fluctuations may be due to the more slowly relaxing gelling systems which implies that fewer independent configurations are being sampled.
The partial structure factors computed for longer copolymers, $N_p=60$ and $N_m=300$, confirm that at equilibrium the core size and shape do not depend on chain length.
  
\subsection{High substitution ratio, $P_{sub}=0.5$}

\begin{figure}[h!]
\centering
 \includegraphics[width=8.3cm]{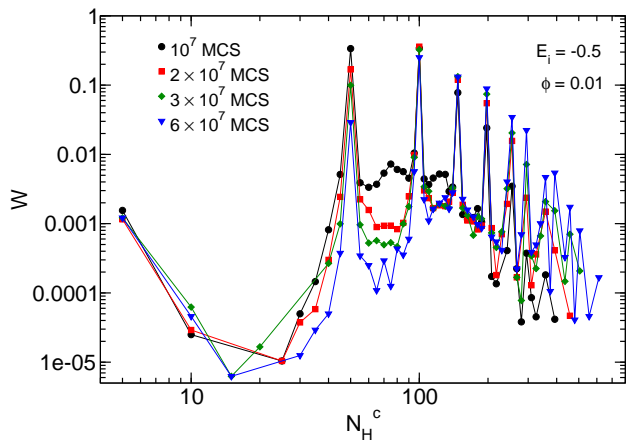}
\caption{Time evolution of the size distribution of the hydrophobic cores, $W(N_H^c)$, for high substitution rate, $P_{sub}=0.5$, for $\phi=0.01$, $E_i=-0.5$, $P_{sub}=0.5$, $B_H=B_P=5$, $N_p=300$, $N_m=100$. The different curves correspond to different simulation times. Note the pronounced peak structure corresponding to exact multiples of the total number of H monomers in a copolymer.} 
\label{fig:time_evol_core_lowc}
\end{figure}

\begin{figure}[h!]
\centering
 \includegraphics[width=8.3cm]{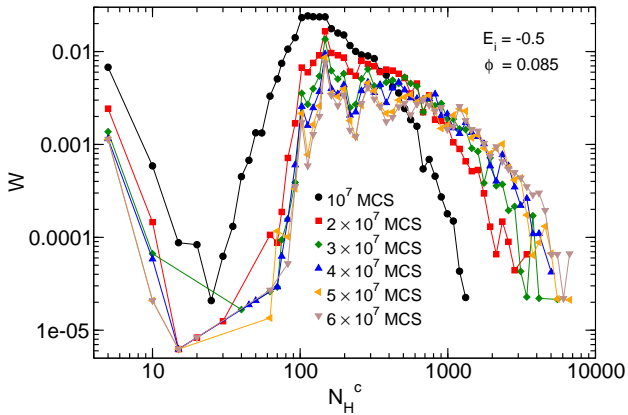}
\caption{Same as Fig.~\ref{fig:time_evol_core_lowc} for $\phi=0.085$. A series of peaks corresponding to exact multiples of the number of H monomers in a polymer develops with time and is most pronounced for intermediate core sizes.} 
\label{fig:time_evol_core_highc}
\end{figure}

Probing systems with a higher substitution ratio allows a different micellar regime to be explored. While for $B_H=4$ and $B_P=16$ well defined spherical micelles are stable in the energy range considered, for $B_H=B_P=5$ they are not. Figures \ref{fig:time_evol_core_lowc} and \ref{fig:time_evol_core_highc} show the time evolution of the size distribution of the hydrophobic cores for $B_H=B_P=5$. Data is presented for strongly interacting H monomers, deep inside the binodal domain ($E_i=-0.5$), and for two concentrations, one below and one above the percolation threshold ($\phi=0.01$ and $0.085$). The size distributions at $\phi=0.01$ and $0.085$ look very different, nevertheless they have some common features, notably the presence of well defined peaks. These peaks appear for monomer numbers that are multiples of fifty, which is the total number of H monomers in a polymer. This indicates that most chains collapse to have all their H monomers in a single core, forming either a monomolecular micelle (size = 50 monomers) or multimolecular micelles (multiples of 50). 

In the low concentration case (see Fig.~\ref{fig:time_evol_core_lowc}) most cores have, at all times, a size which peaks sharply for multiples of fifty. Furthermore, the core sizes markedly evolve towards larger values with time. This is visible in the figure and is also confirmed from the distribution of $N_H^{p,c}$, the number of H monomers in a given core and a given polymer (not shown): at all times the monomers of a given polymer belong to the same core and the cores evolve by aggregation of collapsed polymers. The initial collapse of a copolymer into a globule takes place on a much shorter time scale than the coarsening of the cores. 

At higher concentration, $\phi=0.085$, at early times the core sizes are not discretized in units of fifty, indicating that the initial phase separation is no longer due to the collapse of individual copolymers (Fig.~\ref{fig:time_evol_core_highc}). But with increasing simulation time not only larger cores form as shown by the overall shift of the distribution towards higher values, but also peaks start to appear progressively. The peaks first appear between $10^7$ MCS and $2.10^7$ MCS and then shift to larger multiples of fifty. This is again related to the appearance of clusters of fully collapsed polymers : starting from $2.10^7$ MCS, the peaks of the distribution for intermediate core sizes, corresponding to a few isolated copolymers, develop. This is the same behaviour as for $\phi=0.01$. By contrast, for larger core sizes the peaks tend to be more smeared out. These results suggest that the initial core formation is constrained by the connectivity of the chains, then there is partial reorganisation to form copolymer globules and finally these globules further evolve into larger structures. For a very dense system with $\phi=0.24$ at $E_i=-0.5$, there is still a broad initial maximum and an evolution towards larger micelles, but at no time any sharp peaks appear (not shown).

\begin{figure}[h!]
\centering
 \includegraphics[width=8.3cm]{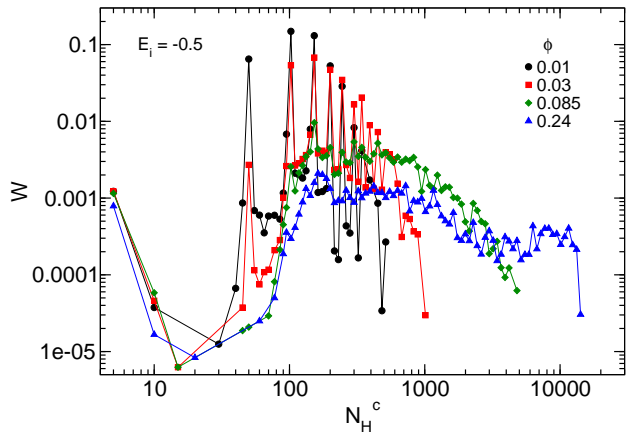}
 \caption{Size distribution of the hydrophobic cores, $W(N_H^c)$, for different concentrations at $E_i=-0.5$, $P_{sub}=0.5$, $B_H=B_P=5$, $N_p=300$, $N_m=100$.}
\label{fig:core_size_vs_phi_psub05}
\end{figure}

The way how the distribution of core sizes is affected by concentration is summarized in Fig.~\ref{fig:core_size_vs_phi_psub05}. With increasing concentration we observe a systematic decrease of the intensity of the peaks and a shift towards larger core sizes. For $\phi=0.01$ the largest core contains up to 500 H monomers (10 chains or more) while for $\phi=0.24$ the largest cores may contain up to 10000 H monomers (200 chains or more). The distribution of the number of chains participating in a core, $W(N_{ch}^c)$, shows a significant shift towards a larger number of chains as well as a broadening when increasing the concentration from $\phi=0.01$ to $0.24$. For $\phi=0.24$ the character of the distribution differs qualitatively from the lower concentrations: there appears a second maximum at very large core sizes typical of the presence of a percolating core structure. Visual inspection of typical configurations confirm the existence of a spanning network of branched cylindrical micelle cores.
\begin{figure}[h!]
\centering
\includegraphics[width=8.3cm]{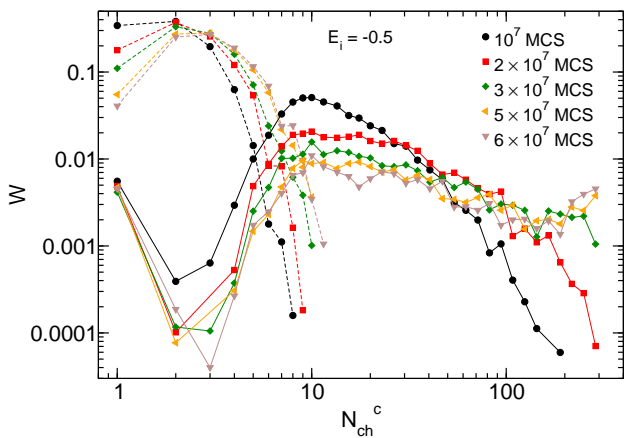}
\caption{Time evolution of the distribution of the number of different copolymers in a core, $W(N_{ch}^c)$, for low ($\phi=0.01$, dashed lines) and for high ($\phi=0.24$, full lines) concentration. Both distributions still evolve. Note the appearance of a second peak for large sizes at $\phi=0.24$.}
\label{fig:core_chain_psub05}
\end{figure}

Fig.~\ref{fig:core_chain_psub05} shows that the time evolution of the distribution at low and at high concentration is totally different. For $\phi=0.01$ a well-defined distribution develops with a maximum around 3 chains per core. For $\phi=0.24$ a very broad peak around eight copolymers per core develops early on and systematically spreads out to larger values at later times while the number of single H blocks and small copolymer cores does not change much. The most significant feature of these curves is the appearance beyond $2.10^7$ of a secondary peak at large sizes, confirming the appearence of a percolating core structure, as already found in Fig. \ref{fig:core_size_vs_phi_psub05}.

The coarsening of the hydrophobic cores is also visible in the evolution of the average core size (not shown). Initially, there is a rapid saturation at the core size corresponding to isolated spherical micelles. As the relaxation is not complete a gradual change corresponding to the reorganisation of the micelles is visible at all concentrations. The onset of this slow coarsening increases with increasing concentration. A similar relaxation behaviour is observed when monitoring the evolution of the average of $N_H^{p,c}$, the number of H monomers in a polymer that belong to a given core.

\begin{figure}[h!]
\centering
\includegraphics[width=8.3cm]{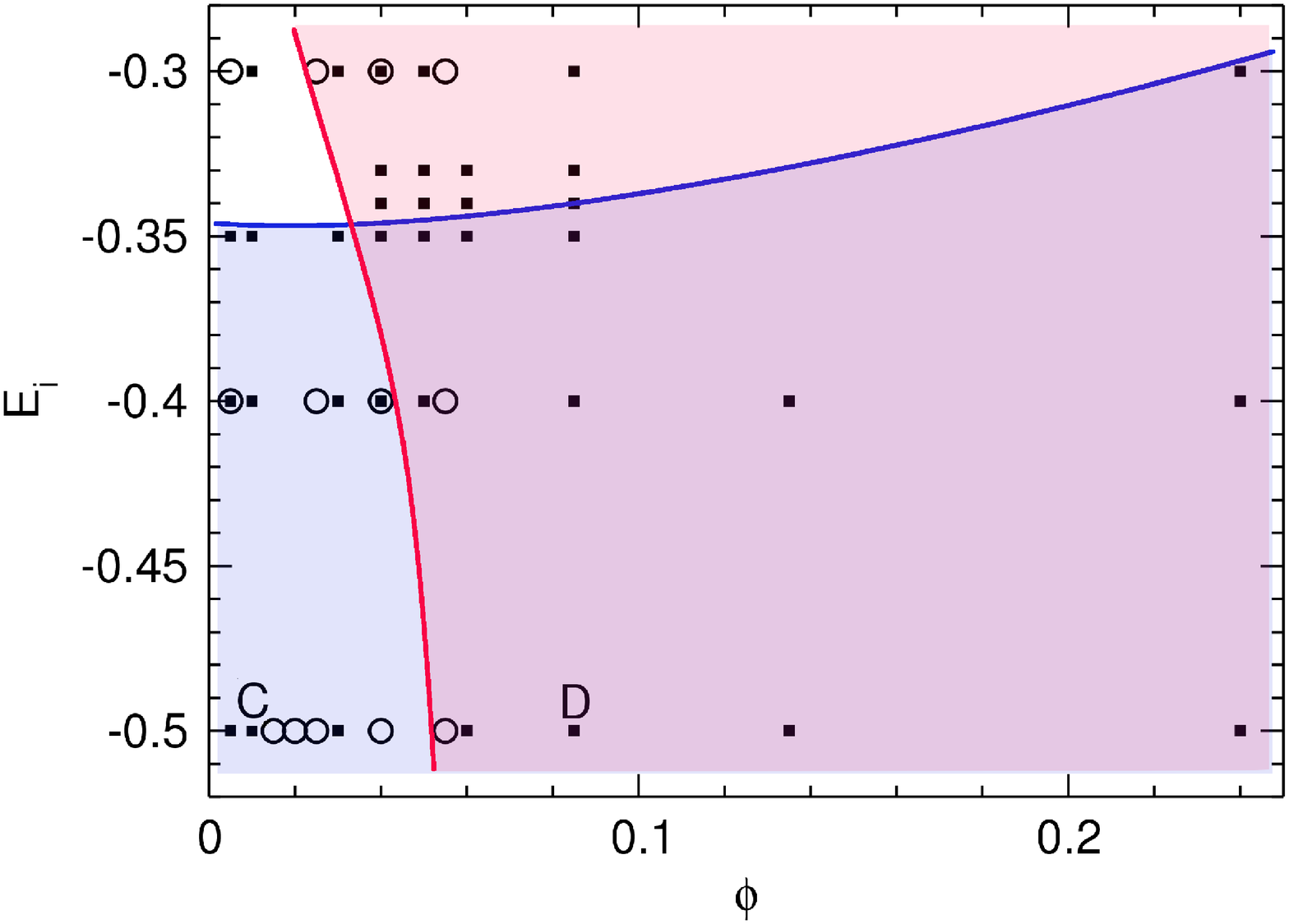}
 \caption{Phase diagram for $P_{sub}=0.5$, $B_H=5$, $B_P=5$. The two data sets are $N_p=50$, $N_m=100$ with $L$ ranging from $L=46$ to $L=100$ (circles) and $N_p=300$, $N_m=100$ with $L$ between $L=50$ and $L=180$ (squares) (same notation as in Fig.~\ref{fig:diag_psub02}). The points for different box sizes coincide, showing that finite size effects are negligible. Snapshots of typical configurations corresponding to points C and D are shown in Fig. \ref{fig:snap_psub05}.}
 \label{fig:diag_psub05}
\end{figure}

\begin{figure}[h!]
\centering
 \includegraphics[width=8.3cm]{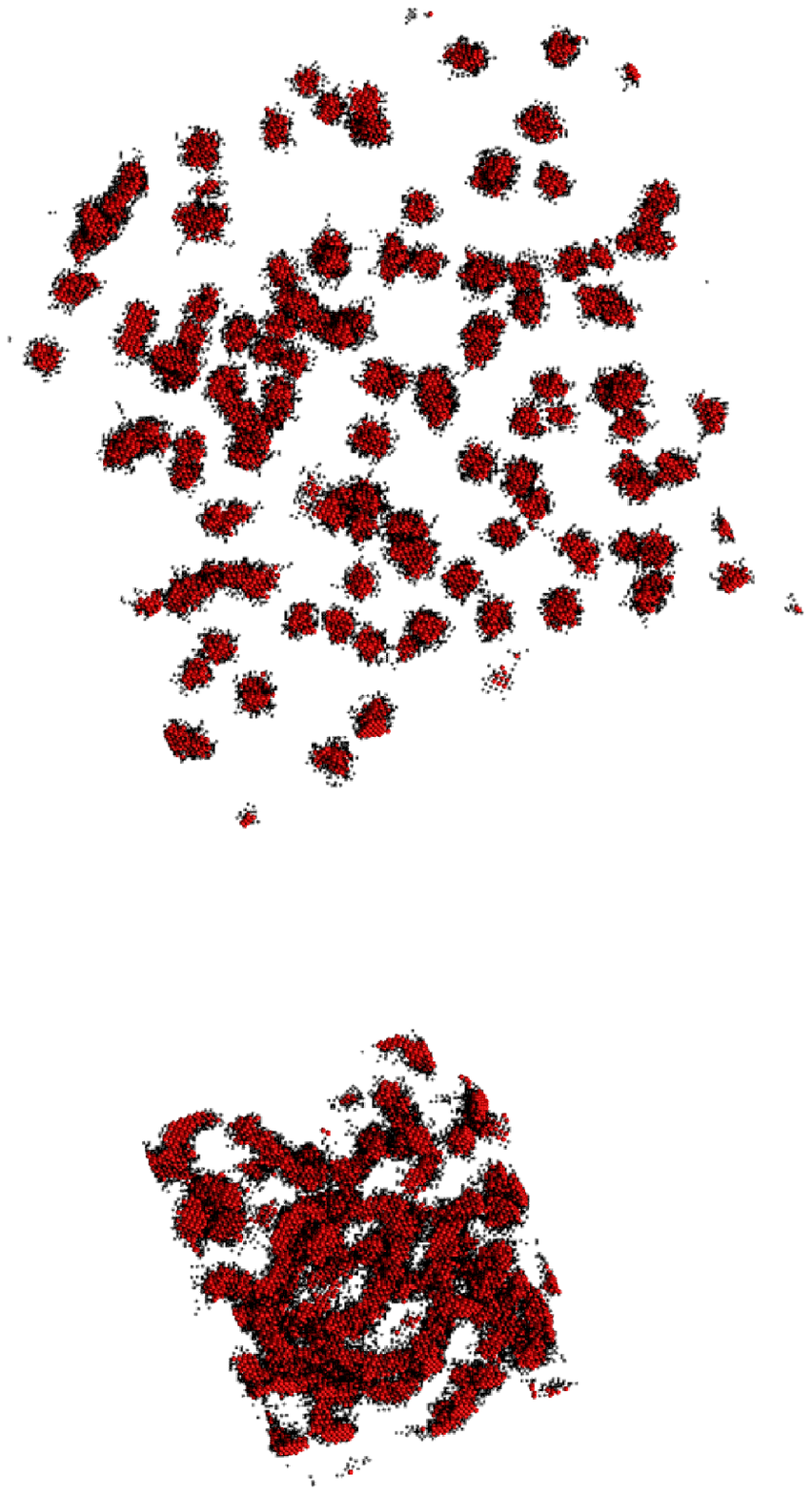}
 \caption{Typical conformations for $P_{sub}=0.5$. Top: simulation of point C in Fig.~\ref{fig:diag_psub05}. Bottom: same for point D, both after $6.10^7$ MCS (same notation as in Fig.~\ref{fig:snapshot_psub02}).}
 \label{fig:snap_psub05}
\end{figure}

The phase diagram for the substitution rate $P_{sub}=0.5$ is presented in Fig.~\ref{fig:diag_psub05} for concentrations ranging from $\phi=0.005$ to $0.24$ and energies ranging from $-0.3$ to $-0.5$. Let us first consider the upper part of the diagram where no micelles are formed. It comprises a range of concentrations above $\phi=0.02$ to $0.03$ where the system percolates. Percolation being defined geometrically and the interaction strength being weak, the gelation criterion may not be meaningful in this domain. Below the line of micelle formation the percolation threshold separates strongly interacting non-gelling systems at low concentrations from gels at high concentrations. The gelation threshold increases with increasing interaction strength.
Snapshots of typical conformations of the system, at the strongly interacting points C and D in the diagram, are shown in Fig.~\ref{fig:snap_psub05}. At low $\phi$, point C, core-shell micelle-like structures are formed, typically spherical or elongated and occasionally connected by hydrophilic blocks. At high concentration, point D, extended linear and/or branched tubular core morphologies are observed. They can be assimilated to wormlike micelles\cite{cates1990statics}. 
Two system sizes with a different number of chains ($N_p=300$ and $N_p=50$) were compared to check for finite size effects. The phase diagram determined from both systems is the same which confirms that the results are not affected by system size.

\begin{figure}[h!]
\centering
 \includegraphics[width=8.3cm]{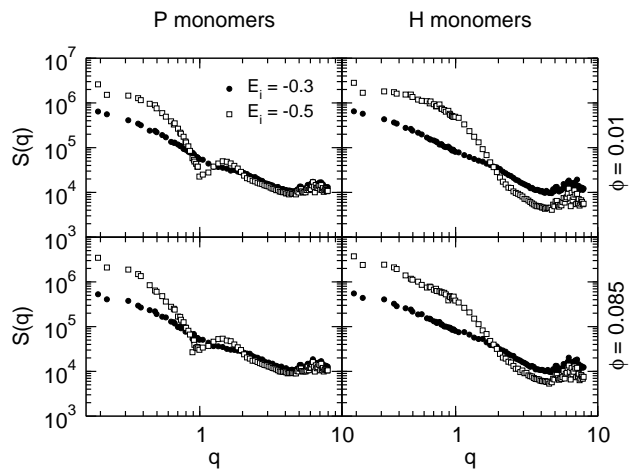}
 \caption{Partial structure factors $S(q)$ for $P_{sub}=0.5$. Top: $\phi=0.01$, bottom: $\phi=0.085$. Left: P monomers only, right: H monomers only.}
 \label{fig:sq_psub05}
\end{figure}

The signature of the structural features are also visible in the partial structure factors presented in Fig.~\ref{fig:sq_psub05}. At low concentration ($\phi=0.01$) $S_H(q)$ behaves, for all $q$, in a similar way as $S_H(q)$ at $P_{sub}=0.2$ and $\phi=0.01$. This is because in both cases the H monomers form spherical micelle cores. At $\phi=0.085$ the shape of $S_H(q)$ is also similar to the $P_{sub}=0.2$ case for large $q$. For low $q$ values it exhibits a $q^{-1.57}$ behaviour instead of the Guinier shape observed for the $\phi=0.01$ cases for $P_{sub}=0.2$ and $P_{sub}=0.5$. This suggests that on a large scale the H cores effectively scale like a polymer in good solvent (this is not strictly true as the H cores are more rigid and ocasionally branch). The hydrophilic partial structure factor $S_P(q)$ is, both for low and for high concentration, in agreement with the expected shell structure, with the minor difference that for $\phi=0.085$ the profile is flatter for small $q$ values. The time evolution of the partial structure factors (not shown) indicates that on intermediate $q$ scales the final conformation (the core-shell structure) develops rapidly. Aggregation then proceeds more slowly on large length scales (low $q$ values), as a result of the ongoing association and reorganization of stable collapsed copolymers.

\section{Discussion} \label{sec:discussion}

The purpose of this study is the investigation of the different scenarios of semidilute solutions of amphiphilic linear multiblock copolymers. For simplicity regularly alternating hydrophobic and hydrophilic monomers were considered. Depending on the hydrophobic substitution ratio $P_{sub}$, the concentration $\phi$ and the strength $E_i$ of the attractive force between the hydrophobic monomers, different types of structures and phases are found, including good solvent behaviour, micelle formation, gelation and phase separation, as summarized in the phase diagrams (see Figs. \ref{fig:diag_psub02} and \ref{fig:diag_psub05} for different substitution ratios). 

For low concentration and weak interaction between the H monomers (white area in the diagrams) the system forms neither micelles nor a gel for any substitution ratio. With increasing concentration there is a sol-gel transition that is no different from gelation of homopolymers.

When increasing the interaction strength while staying at low concentrations self-assembly takes place (the area below the blue line in the phase diagrams). A common feature of the self-assembled structures is the formation of cores of hydrophobic monomers surrounded by shells of hydrophilic monomers, as a result of the short-range attractive interaction between the H monomers. Core formation of the copolymers may be triggered either by increasing the interaction energy or the concentration. The threshold for structure formation depends on the hydrophobic substitution rate and on the block lengths, as expected from dilute solutions\cite{hugouvieux_amphiphilic_2009}: the lower the entropic contribution of the P monomers (the shorter the P blocks in the chain), the lower the attractive interaction strength needed for triggering the association of the H monomers. 

Depending on the hydrophobic substitution rate and the chain length, copolymers show different types of core-shell structures, ranging from spherical micelles and pearl-necklaces of spherical micelles to linear and branched tubular micelles. It is influenced by the P layers surrounding the H cores and the connectivity between the H and P moieties. The protection of a hydrophobic core depends directly on the lengths of the H and P blocks: Decreasing the number of hydrophilic monomers in the chain leads to a thinner P shell as shown by the more pronounced intermediate peak in $S(q)$ of the P monomers (compare Figures \ref{fig:sq_psub02} and \ref{fig:sq_psub05}).

For thick P shells ($P_{sub}=0.2$ in our study), the phase separation of the H monomers into cores is stopped by the protective P layer as can be seen in Fig. \ref{fig:core_size_vs_phi_psub02}. An increase of the core size would require P monomers to be included in the core, which is energetically unfavorable, or the entropic penalty from densifying the shell of P monomers would be too strong. Conversely, for large values of $P_{sub}$ ($P_{sub}=0.5$ in our study), the rather thin P shell does not protect fully against further association of the H blocks. This lack of protection leads to elongated core-shell structures, elliptical at low concentrations and tubular and/or branched at higher concentration. Put differently, the thickness of the outer P layer determines which micellar structure stabilizes. 
A result similar to our simulations has been found experimentally\cite{determan_supramolecular_2006, determan_ph-_2008} where spherical core-shell structures become elongated and eventually precipitate when increasing the hydrophobicity of the copolymers. 

At $P_{sub}=0.2$, when the micelles are well protected, the phase structure is quite similar to micellization in surfactant or diblock copolymer solutions, where, in a narrow range of $E_i$, increasing concentration induces micelle formation (CMC)\cite{israelachvili_intermolecular_1991}. In all simulations the micelles have a well-defined size distribution which barely evolves with increasing concentration, while the number of micelles increases linearly. Also, the higher the concentration (in the range of $E_i$ between -0.47 and -0.52), the lower the interaction strength needed for micelle formation. This is not surprising in the light of the fact that, for the chain lengths and simulation parameters considered, stable micelles are made of several copolymers. There are, however, noticeable differences with respect to surfactants. For strong interaction energies, copolymer micelles form at any concentration, the CMC is zero. This is because intramolecular micelles form at any polymer dilution as the connectivity locally forces a finite concentration of H monomers. For the same reason, the micelle phase boundary (see Fig.~\ref{fig:diag_psub02} and Fig.~\ref{fig:sq_psub05}) is rather flat.

A parallel can be made between the effects of $E_i$ and $P_{sub}$. Increasing the interaction between H monomers at low values of $P_{sub}$ where pearl-necklaces of micelles are present leads to the formation of tubular structures with the same radius. Indeed, for a given radius imposed by the block structure, the tube is energetically more favorable because of its lower surface to volume ratio. Upon increasing the interaction strength, this energetic gain eventually overcomes the entropic penalty of accommodating the P monomers around a tube. On the other hand, starting from a tubular structure and increasing the length of the P blocks (decreasing the ratio of hydrophobic monomers in the chain) at constant $E_i$ would lead to the break-up of the tube into several cross-linked spherical micelles, a pearl-necklace.

The connectivity of the P and H moieties has a strong influence on the core-shell structures. For low substitution rates ($P_{sub}=0.2$) the length of the H blocks determines the radius of the H cores, in the case of highly substituted copolymers ($P_{sub}=0.5$), the length of the H blocks limits the diameter of the tubular micelles. This relationship has been shown to hold for long copolymers in dilute solutions\cite{hugouvieux_amphiphilic_2009}. It also works quite well for more concentrated solutions: for $P_{sub}=0.2$ and $E_i=-0.55$ the core diameter estimated from $S(q)$ for the H monomers is 3.8, to be compared with the block length 4; for $P_{sub}=0.5$ and $E_i=-0.5$ the tube diameter is 5.2, for a block length of 5.

The present simulations explore concentration effects due to aggregation of collapsed globular copolymers. At low substitution rates ($P_{sub}=0.2$) and low concentrations, the self-assembled structures are shown to be either multimolecular ($(H_4P_{16})_5$, $N_m=100$) or approximately monomolecular ($(H_4P_{16})_{15}$, $N_m=300$). From the comparison of the two cases one concludes that the connectivity has little influence on the micelle stabilisation. This result is consistent with the findings of Gindy \textit{et al.}, whose simulations showed that increasing the number of diblock motifs in copolymer chains leads to a decrease in the most probable number of chains involved in a core\cite{gindy_phase_2008}. It was also shown experimentally by \citet{zhou_thermo-induced_2007} that for a constant number of diblock motifs in the chain, increasing the length of both blocks while keeping the hydrophilic/hydrophobic ratio constant induces a transition from multimolecular micelles to monomolecular micelles. Therefore the number of H monomers needed for the formation of stable core-shell structures is determined primarily by the energy-entropy balance and not by the chain length or the block structure.

At low concentration and for the thermodynamic parameters and the chain lengths considered in our simulations, the H blocks of a copolymer all collapse into a single micelle core. The structural changes upon increasing concentration is very different for low and for high $P_{sub}$. For $P_{sub}=0.2$ the core-shell structure does not change significantly when increasing $\phi$ but the H blocks of a given copolymer are now shared between several cores thus providing hydrophilic links between different cores (this is particularly prominent for $N_m=300$ where the cores are monomolecular at low $\phi$ and intermolecular at high $\phi$). At $P_{sub}=0.5$ the core structure is not preserved when increasing $\phi$ because the low concentration spherical/elliptical micellar cores aggregate as a whole to form larger cores.

% Gelation
For gelation we use the static percolation threshold as the criterion. For weak energies, \textit{i.e.} $|E_i|<0.48$ for $P_{sub}=0.2$ and $|E_i|<0.35$ for $P_{sub}=0.5$, when no micelles are formed, percolation occurs for purely geometric reasons and does not necessarily lead to a physical gelation transition. For energies where phase separation of the H monomers sets in, percolation does correspond to gelation. At $P_{sub}=0.2$ gelation is observed either by increasing concentration or upon strengthening of the interactions. A particularity of the concentration driven transition is that gelation not only consists of connecting nearby micelles, but also because  increasing the concentration induces the hydrophobic blocks of a copolymer to spread over several micelle cores which effectively increases size and entanglement of the copolymers. The gel looks rather weak as it consists of hydrophilic cross-links between spherical hydrophobic cores. Interestingly, in a narrow concentration range the system undergoes a reverse gel-sol transition at quite strong interactions. This transition is due to the copolymer contraction upon strengthening the interaction, as directly visible from the chain radii which decreases from $R_g^2\approx 23$ at $E_i=-0.54$ to $R_g^2\approx 17$ at $E_i=-0.6$.
For $P_{sub}=0.5$ the percolating network is very different: The spanning cluster is made of a continuous structure of hydrophobic monomers, covered by a shell of hydrophilic monomers. The evolution of the system from spherical micelles to tubular structures to a spanning network proceeds by coalescence since at high density the P shell cannot prevent the H monomers of different cores from interacting. Furthermore, the tubes tend to be straight because the P shell at the tube ends is less dense which favors growth at the ends. Also, at $P_{sub}=0.5$ no reverse transition from gel to sol occurs at strong energies because increasing the energy does not significantly affect the topology of the hydrophobic tubular cores. 
Consistently with the different cases for $P_{sub}=0.2$ and $0.5$, simulations of pentablock copolymers in solution by Anderson and Travesset\cite{anderson_coarse-grained_2006} have provided evidence for what they call a swollen gel (a network of loosely connected micelle cores) and for a giant cylindrical micelle (a single tubular core made of several copolymers) when increasing the ratio of hydrophobic to hydrophilic monomers in the chain. Also, \citet{gindy_phase_2008} observed in simulations, at $P_{sub}=0.5$, the formation of a different kind of gel network for what they call microscopically or macroscopically phase separating copolymer, depending on the lengths of the H blocks, the P blocks and the copolymers.

The present simulations distinguish, depending on the parameter values, between two classes of micelles, well protected ones ($P_{sub}=0.2$, spherical) and poorly protected ones ($P_{sub}=0.5$, tubular). For the chain lengths considered each micelle consists of one or a few fully collapsed copolymers. Our low concentration calculations\cite{hugouvieux_amphiphilic_2009} for much longer chains ($N_m > 300$) show that under well protected conditions pearl necklace structures and under poorly protected conditions extended tubular (wormlike) structures appear. Extensive calculations for these cases at high concentration are computationally demanding but one may guess the gel structure that develops in this case upon increasing concentration. For pearl necklaces occasional bonding of micellar pearls, by the same mechanism as the hydrophilic bonding of spherical micelles in our $P_{sub}=0.2$ simulations in Fig.~\ref{fig:snapshot_psub02}, will lead to a loosely and hydrophilically connected network of necklace chains. For the tubular case the aggregation process is no different from the $P_{sub}=0.5$ one in Fig.~\ref{fig:snap_psub05} but with a coarser network structure because with increasing chain length the elementary micelles simply are more elongated.

Regarding the phase behaviour of methylcellulose, the present model gives a possible explanation for the two different types of gels observed experimentally: a weak, transparent gel at low temparature and a strong, turbid gel at higher temperature. The clear gel at relatively low concentrations and weak interaction strength then corresponds to the micellar phase connected by hydrophilic strands and the turbid gel at higher temperature (because of the LCST, higher temperature corresponds to stronger interactions) corresponds to a gel formed by a network of thick hydrophobic strands. In an experimental situation, and particularly for methylcellulose, the distinction between hydrophobic (strongly interacting) and hydrophilic (non-interacting) blocks is of course not as clearcut as in the model, therefore the hydrophilic monomers will also contribute to the structure formation. For methylcellulose the degree of methylation of each glucose cycle ranges from 0 (hydrophilic) to 3, (hydrophobic)\cite{hirrien_thermogelation_1998,kobayashi_thermoreversible_1999,arisz_substituent_1995}. Therefore, in the strong gel regime intermediate degrees of substitution provide some interaction between the strongly interacting hydrophobic cores. The strand formation observed in cryo-TEM images suggests that rather stiff bundles form in the strongly phase separating part of the phase diagram\cite{gottlieb,bodvik}. Thus the multiblock model can explain the main features of methylcellulose phase diagram. The observed bundling \cite{gottlieb} could be included by introducing a weak attraction between the hydrophilic monomers in our model. There is one puzzling experimental feature that is not reproduced in the simulations: in the weak gel region there is a gel to sol transition upon increasing concentration \cite{chevillard_phase_1997}. Possible explanations for this counterintuitive observation could be a more complicated interaction structure (e.g. involving hydrogen bonds) or that the rheologically determined gel-sol transition is not a macroscopic phenomenon but the effect of intermediate scale structural reorganisations. 

\section{Conclusions}

In the present study a generic lattice copolymer model is used to assess the phase behaviour of amphiphilic multiblock copolymers in semidilute solution, as a function of the solvent quality of the H monomers, the length of the hydrophobic and hydrophilic blocks in the chain, and the copolymer concentration. The simulations provide evidence for two distinct types of gel structures, depending on the stability of the micelles that form when the hydrophobic monomers phase separate. For chains of spherical micelles, also called pearl necklaces, a weak, hydrophilically linked gel forms when, with increasing concentration, the micelle cores exchange hydrophobic blocks belonging to different copolymers. Such a gel is a direct consequence of the multiblock nature of the copolymer. For unstable micelles a strong gel builds up when spherical or tubular elementary micelle cores join to form larger linear or branched structures. By lowering the solvent quality one passes from the weak gel to the strong gel phase. This provides a possible explanation for the weak and the strong gel phase observed in methylcellulose.

\section{Acknowledgements}
The authors thank Moshe Gottlieb for several enlightening discussions and for communicationg unpublished experimental work on methylcellulose. Computing support from the CINES in Montpellier is acknowledged.

\footnotesize{
\bibliography{version_16092010} 
\bibliographystyle{rsc} 
}

\end{document}